# Automated experiment in 4D-STEM: exploring emergent physics and structural behaviors

Kevin M. Roccapriore,[1,*] Ondrej Dyck,[1] Mark P. Oxley,[1]
Maxim Ziatdinov,[1,2] and Sergei V. Kalinin[3,*]

[1] Center for Nanophase Materials Sciences, Oak Ridge National Laboratory, Oak Ridge, TN 37831
[2] Computational Sciences and Engineering Division, Oak Ridge National Laboratory, Oak Ridge, TN 37831
[3] Department of Materials Science and Engineering, University of Tennessee, Knoxville TN, 37916

ABSTRACT
Automated experiments in 4D Scanning Transmission Electron Microscopy are implemented for rapid discovery of local structures, symmetry-breaking distortions, and internal electric and magnetic fields in complex materials. Deep kernel learning enables active learning of the relationship between local structure and a 4D-STEM based descriptors. With this, efficient and 'intelligent' probing of dissimilar structural elements to discover desired physical functionality is made possible. This approach allows effective navigation of the sample in an automated fashion guided by either a pre-determined physical phenomenon, such as strongest electric field magnitude, or in an exploratory fashion. We verify the approach first on pre-acquired 4D-STEM data, and further implement it experimentally on an operational STEM. The experimental discovery workflow is demonstrated using graphene, and subsequently extended towards a lesser-known layered 2D van der Waal material, $MnPS_3$. This approach establishes a paradigm for physics-driven automated 4D-STEM experiments that enable probing the physics of strongly correlated systems and quantum materials and devices, as well as exploration of beam sensitive materials.

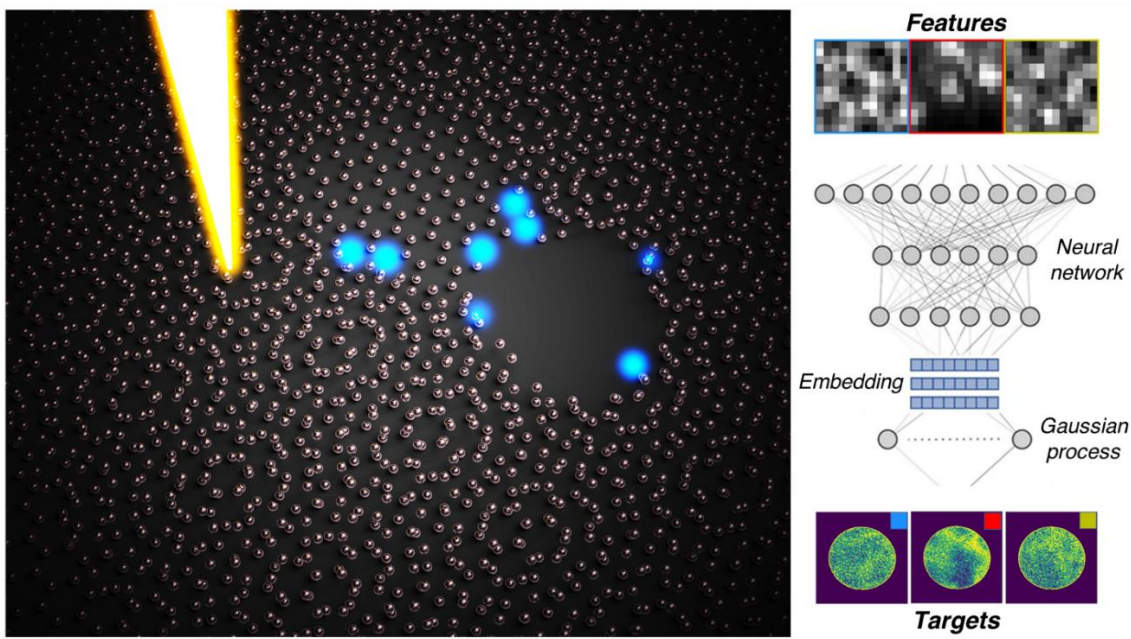

The gamut of functional properties of materials is indelibly linked to their atomic structure. The chemical reactivity and mechanical properties alike are linked to the details of the atomic bonding and connectivity of the atomic network. The catalytic and electrocatalytic behaviors are tied to the structural and electronic properties of a group of atoms within a host material. Emergent functionalities of strongly correlated and quantum materials associated with minute changes in electronic structure and coupling between electronic and spin systems are often reflected in symmetry breaking of high-symmetry crystallographic lattices.

This inherent coupling between chemical and physical phenomena and the structure and structural distortions renders structural imaging techniques such as (Scanning) Transmission Electron Microscopy ((S)TEM) an inherently powerful tool to probe these on the atomic level. The pioneering work Jia[1] has demonstrated the potential for mapping ferroelectric polarization fields, launching a decade of focused research effort.[2-5] This capability was further extended towards the probing the octahedra tilts,[6,7] emergence of topological defects,[8,9] nanoscale phase separation in perovskites and charge-density wave formation in layered dichalcogenides and other systems.[10-12]

Recently, the introduction of fast digital cameras and computational data analytics enabled rapid progress in 4D-STEM, a technique based on acquisition of local diffraction patterns. In this method, the size of the excitation volume is significantly smaller than the distance between the scatterers, giving rise to highly unusual diffraction physics. Originally, 4D-STEM was proposed by Rodenburg as a pathway to avoid intrinsic resolution limits in STEM.[13,14] Using this approach, atomic resolution imaging has been demonstrated along with imaging of the light elements.[15] Similarly, 4D-STEM provides insights into the structure of electric and magnetic fields on the atomic level, allowing local chemistry analysis of defects,[16-19] mapping of lattice distortion and interlayer spacing in 2D materials,[20] and even visualization of anionic electrons.[21]

However, as with any electron beam method, 4D-STEM is associated with a significant number of limitations. The data acquisition times for the local diffraction mapping and hence associated electron doses are fairly high, limiting its applications for beam-sensitive materials. Similarly, for many materials systems the behaviors of interest are concentrated in a small number of spatial locations. In 2D materials, features of interest are dopant atoms and point or extended defects. In 3D materials, these are often concentrated at the interfaces between dissimilar materials, topological defects in systems with multiple variants or other forms of symmetry breaking phenomena, or nanoscale phase separation in systems with structurally degenerate ground states. Here, the grid sampling approach that collects information from all points equally is suboptimal for many reasons. Aside from being computationally expensive and imparting unnecessarily large amounts of electron dose to the specimen, the sample space is far too large to sufficiently probe for answers to the physics questions in mind. This necessitates the development of approaches allowing the collection of information from regions of interest and combining exploration of the novel physical phenomena and exploitation towards the known ones.

Here we introduce an approach for structural discovery in 4D-STEM using deep kernel learning (DKL),[22,23] which is illustrated in **Figure 1**. In the exploratory mode of the DKL experiment, the algorithm builds the relationship between local structure visualized via the dark-field image and the 4D-STEM diffraction pattern. We further extend this approach for the automated experiment driven by physical discovery of parameters that can be extracted from local diffraction including relative changes in electric field strength, charge density, and lattice strain.[24-26] This is illustrated for twisted bilayer graphene and the recently-discovered periodic beam-induced patterns in the van der Waals 2D material, $MnPS_3$.[27]

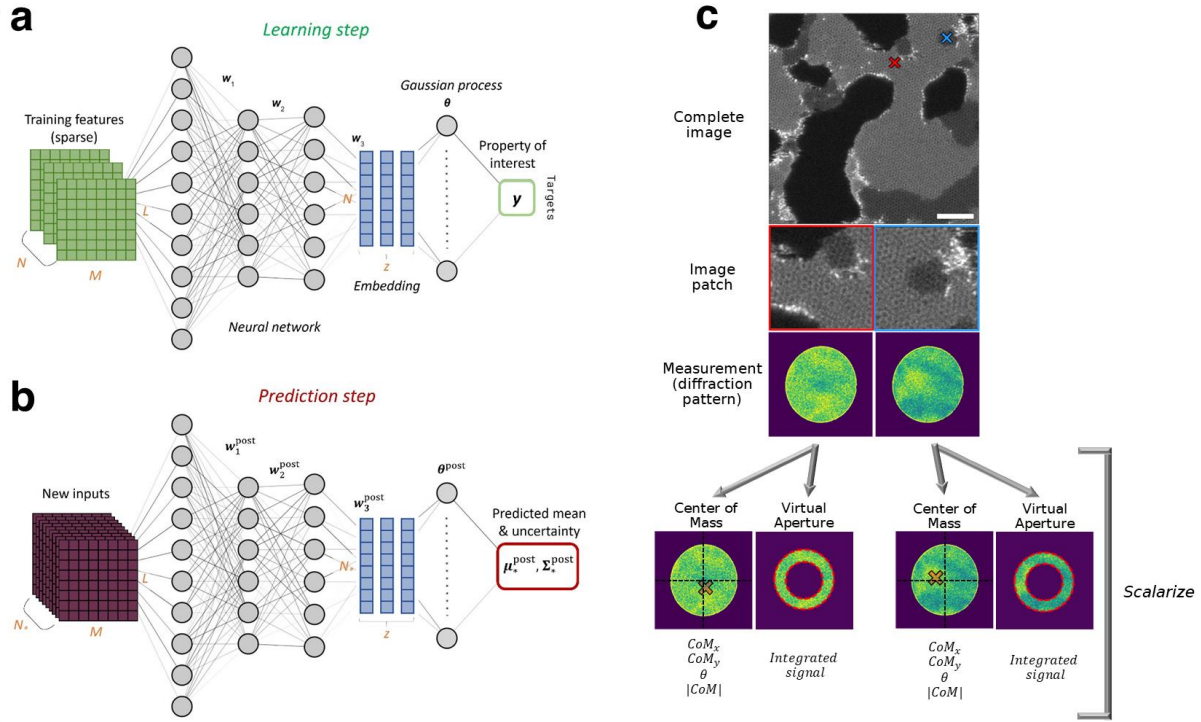

**Figure 1**. Deep kernel learning schematics and workflow for 4D-STEM. Each iteration consists of a learning step, **a,** and prediction step, **b**, where measurements are collected as according to **c**. "Features" are selected image patches from the HAADF-STEM image, while "targets" are the corresponding (scalarized) diffraction patterns acquired from the center coordinate of the image patches. Measurements of diffraction patterns are acquired and scalarized according to the property of interest to be explored as it relates to local structure from image patches, illustrated in **c**. Many quantities can be calculated from 2D diffraction pattern; center of mass and virtual apertures transformations are shown, from which individual components are then derived.

Here, we briefly discuss the principles of the DKL workflow. Similar to many other active learning methods, DKL is based on the concept of Gaussian Processes (GP).[28] GP in turn refers to a broad class of methods of functional interpolation, i.e., reconstructing the (unknown) function $f(x)$ defined over parameter space $x \in \mathbf{R}^n$. Here, it is assumed that the measurements $y_i$ at points $x_i$ yield the noisy function values, $y_i = f(x_i) + \varepsilon$, where $\varepsilon$ is a normally-distributed observation noise. In the context of microscopy, the parameter space $x$ corresponds to the image plane, $x = (i, j)$ and $y$ is a property measured in the $(i, j)$ coordinate. However, the same concept can be applied universally over other, sufficiently low-dimensional, parameter spaces, e.g., compositional phase diagrams in synthesis, controls for the aberrations in microscope tuning, etc.

The reconstruction in GP is based on the concept of the kernel function, describing the strength of correlations between the adjacent regions in the parameter space. The functional form of the kernel is defined prior to the experiment, whereas kernel parameters such as the associated length scale are learned from measured experimental data. The GP regression subsequently yields the expected function values, $\mu_*^{post}\mu$, over the yet unmeasured points of the parameter space and the associated uncertainties, $\Sigma_*^{post}$. In the purely exploratory mode, the location for the next measurement is chosen such as to minimize the uncertainty, $x_{N+1} = argmax(\Sigma_*^{post})$. In the Bayesian optimization (BO), the prediction and uncertainty are combined in the acquisition function, $acq(\mu_*^{post}, \Sigma_*^{post})$, the maximum of which guides the navigation. This allows one to combine the exploration and exploitation of parameter space towards required behaviors.

Despite its utility in many control and optimization tasks, the applicability of pure BO in imaging is limited. This happens because the real-space images often possess complex hierarchical structures, large regions with constant contrast, and sharp interfaces that cannot be described by the GP's kernel function. The alternative solution is offered by deep kernel learning (DKL), combining the representation power of deep neural networks and the flexibility of the GP[29] (Fig. 1a, b). In DKL, a neural network is used to embed data into a low-dimensional latent space where a GP kernel operates. The weights of the neural network and the kernel hyperparameters are learned simultaneously via a gradient ascent on marginal log-likelihood. In the context of microscopy, DKL seeks to establish the relationship between the small patches of the structural image and the corresponding functional responses (Fig. 1c), similar to the recently introduced *im2spec* approaches.[30,31] Here, functional responses are the properties that can be calculated from the 4D-STEM data, e.g., electric fields, strain, etc. The *im2spec* approaches are based on the supervised deep learning and requires the availability of the full grid spectroscopic measurements. However, in DKL the correlative relationship is established in the active learning mode from available (sparse) measurements and updated once the new data becomes available. The exploration-exploitation is built based on the acquisition function combining the classical uncertainty part and physics-based quantity derived from the predicted quantity. For example, recently we have demonstrated the use of the surface plasmon peak intensity towards the exploration of the edge plasmon in $MnPS_3$ and regions with the enhanced electromechanical response in PZT.[22,23] Throughout this work we have used the "expected improvement" acquisition function (see its definition in Methods). We extend this approach to the 4D-STEM modality, expanding the capabilities of automated experiments in STEM.

RESULTS/DISCUSSION

As a model system, we choose twisted bilayer graphene (TBG). TBG is host to a gamut of interesting physical behaviors, many of which depend on the relative twist angle between the two layers - a part of the field of twistronics.[32-34] For example, the so-called twisted bilayer "magic angle" graphene displays superconductivity as a result of the flat band topology caused from the modification of the potential energy landscape. Edges, holes, and other defects in both graphene and TBG are expected to affect the twist behaviors, and hence potentially control these behaviors locally.[35,36] The electron beam in the STEM, however, can alter or destroy the defects, even below the knock-on damage threshold for graphene due to a lower binding energy at edges or defects or the presence of a second layer in bilayer graphene.[37,38] Consequently with TBG, capturing a 4D-STEM or STEM-EELS dataset causes the sample to often change during the acquisition - in other words, it is highly dose sensitive. Aside from dose sensitivity, collection of high quality 4D data is limited to small, nanometer sized regions; sampling large regions is prohibitive since it is costly to obtain high quality data over large fields of view.

One strategy towards mitigation of damage in beam sensitive specimens is intelligent sampling. DKL accomplishes this because a) between training steps, the specimen is blocked from electron irradiation, and b) a small fraction of the total number of available points are ever visited. In the DKL workflow, the HAADF-STEM image is obtained prior to the intelligent sampling of the 4D data such that the model has complete access to the image space; therefore, there is a small (compared to full 4D data acquisition) but unavoidable dose occurring from this first step. To illustrate the DKL approach to 4D-STEM, we first show two ground truth 4D datasets in **Figure 2** of single and bilayer graphene in which the diffraction pattern is recorded at all pixel positions. These examples contain a variety of defects and dopant atoms that serve to illustrate the DPC method. With each diffraction pattern, the center of mass (CoM) of the central beam is calculated which, by relative CoM shifts, is used to compute the local electric field and related quantities such as charge density and electric potential at the sample. We note that these measurements are *relative* and not intended to be quantitative since the probe itself is convolved with the fields but are nonetheless useful to understand the relationships between structure and derived 4D quantity.

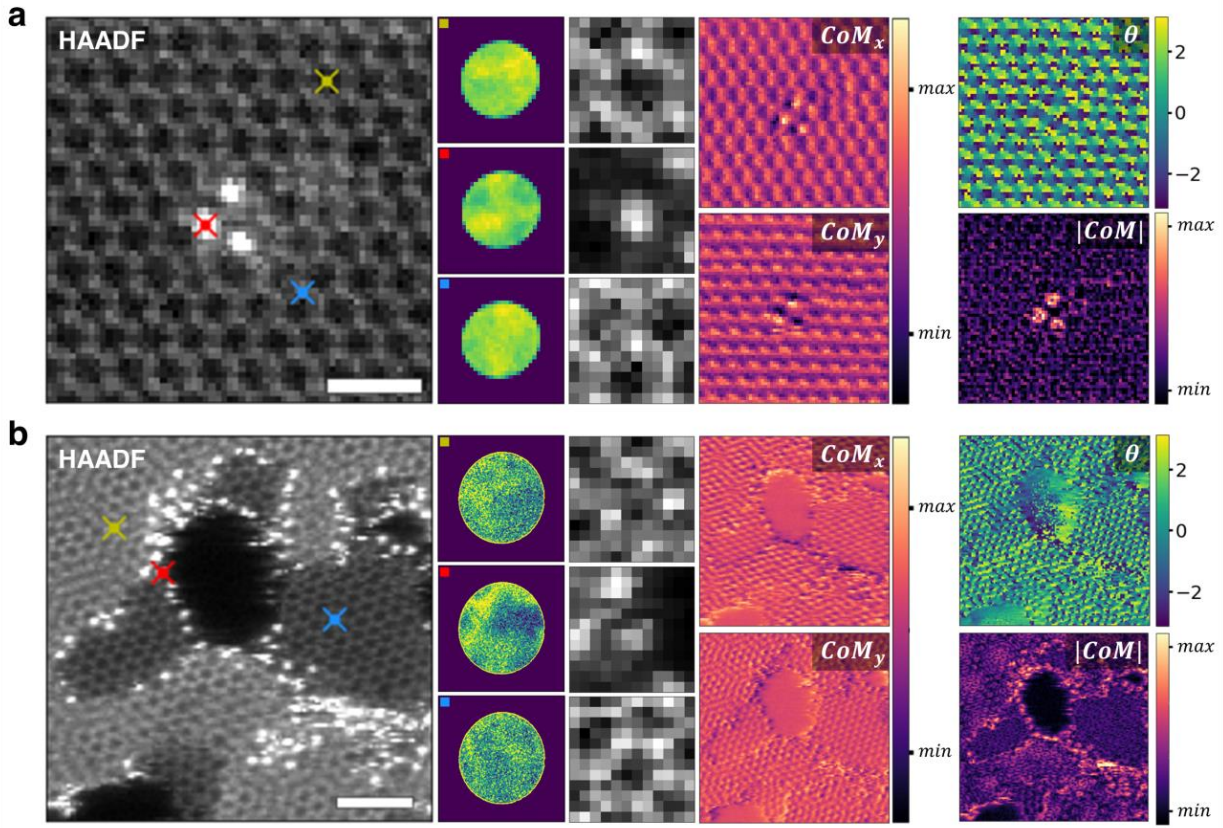

**Figure 2**. Ground truth 4D-STEM of single layer and bilayer graphene, both containing defects. HAADF-STEM images with selected ronchigrams and their local image patches are shown. CBEDs are reduced to their CoM-based scalar quantities: $CoM_x$, $CoM_y$ $angle$, and $magnitude$. HAADF scale bars 0.5 nm and 2 nm, $\theta$ in units of radians.

For an autonomous or automated experiment, each measurement should help inform the subsequent measurement by increasing the model's knowledgebase of the selected structure-property relationship. In 4D-STEM, however, each measurement is a 2D diffraction pattern, therefore the diffraction pattern must be reduced to a single quantity, i.e., it must be scalarized. Here we use the calculation of the CoM that reduces the 2D image to a single vector, whose components can then be used as the measured property. In this way, four quantities can be used – the *x* component, *y* component, angle, and magnitude. However, the scalarization (or defining behavior of interest) can be based on more complex analyses, including physics-based inversions of the 4D-STEM data towards the scattering potential, and selecting associated features of interest. We note that the descriptors derived from 4D STEM diffraction data are linked to a small patch in the structural image space of size *w* centered at the spatial coordinate from where they are acquired. In this way, the image patch size *w* may be varied to explore the connection between the derived functional response and its localization in real space.

The ground truth – which here simply has the meaning of a complete data set that was acquired in a normal grid-based fashion – in **Figure 2** shows diffraction patterns from several selected locations. Some differences in contrast among the local diffraction patterns may be discernible by eye, but in no way is the physical meaning of these differences obvious. The CoM shift, relative to a vacuum reference, is calculated for all spatial positions and the vector components and magnitude are all shown. The most apparent result is that the CoM magnitude is strongest surrounding dopant atoms, which are higher in atomic number and therefore exhibit a stronger atomic electric field and influence on the beam. The AA stacking sites in the bilayer graphene are also recognizable from the CoM magnitude contrast. Motivated by discovering the structures that exhibit the strongest field strengths felt by the beam, the CoM magnitude is most natural to use as a scalarizer for active learning, but others are compared later for completeness.

Application of DKL to the same data sets is shown in **Figure 3** for varying degrees of provided data. Representation of the CoM magnitude can be achieved with very few numbers of measurements, even less than 1% of the total space can provide reasonable predictions with small uncertainties, implying that the relationship between HAADF structure and CoM magnitude is realized rather quickly. With an increasing number of measurements, the primary difference in predictions begins to be deviations in local contrast but the same features are still present – Figure 3(b) illustrates this behavior well when comparing 100% to 10% and 1% measurements. In all cases, DKL has complete access to the full structural information; consequently, local image patches are immediately obtained when sampling the diffraction space. An important consideration here is that the supplied "measurements" are *randomly* chosen, in contrast to a true active learning experiment where future measurements are chosen based on prior measurements. In other words, the points are not optimally chosen to maximize the cumulative knowledge the model acquires; however, this is purely for illustrative purposes as DKL would not select random measurement locations in a true active learning setting.

As an aside, the DKL reconstruction with 100% provided data in Figure 3(a) highlights peculiar behavior that is only weakly observed in other imaging modalities – it is as if there is a smear of contrast near the trimer set, however the only species present are carbon in the graphene network and the silicon dopant atoms – contamination effects are typically obvious. Upon looking closely at the HAADF from Figure 2, this feature is also partially evident, and we postulate it is due to hydrocarbon molecules pinned to the dopants whose movement must be considerably faster than the beam motion and therefore the scattered intensity is smoothed out.

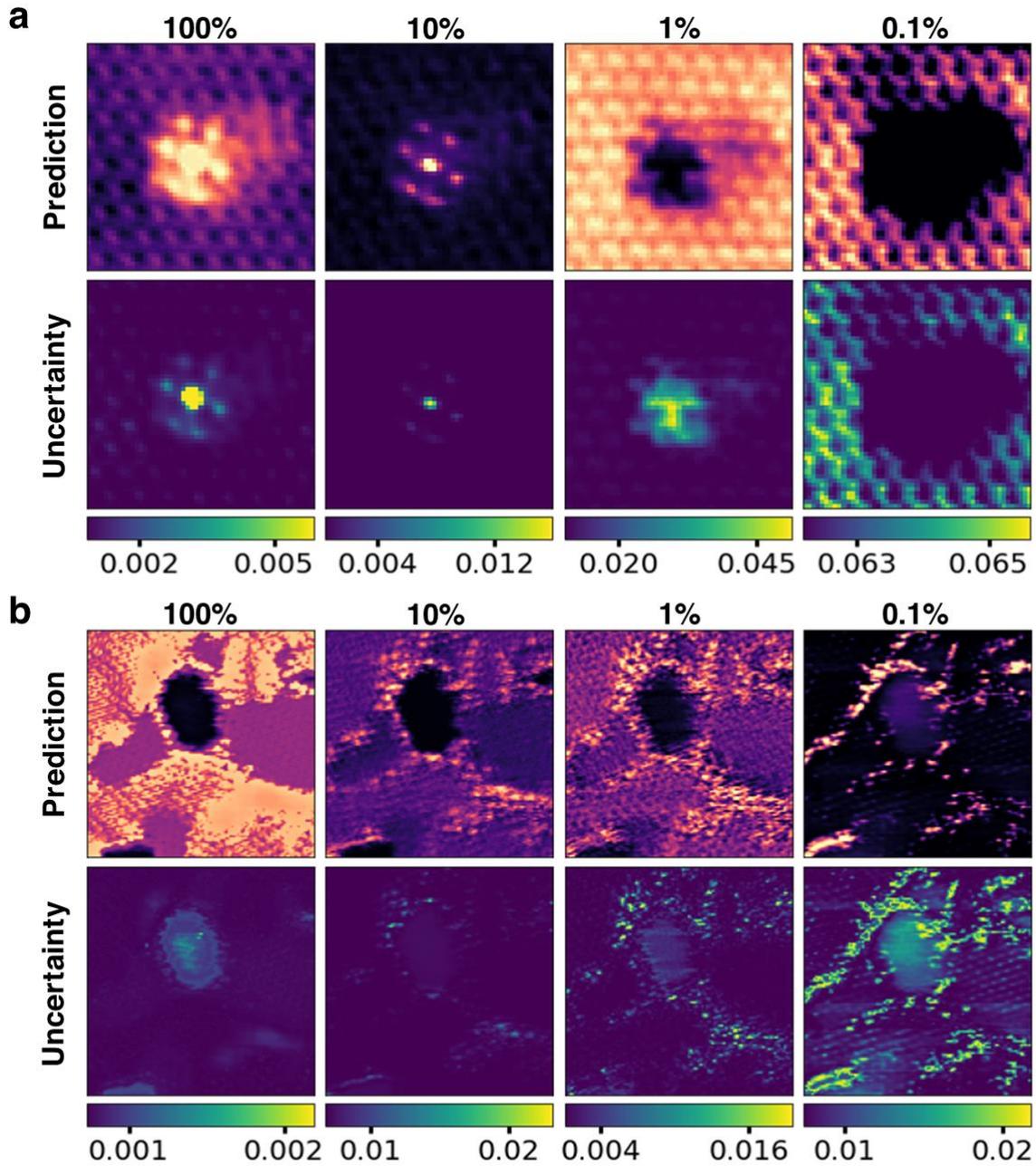

**Figure 3**. DKL reconstruction with CoM magnitude scalarizer for single layer (**a**) and twisted bilayer graphene (**b**). Reconstructions of the predicted CoM magnitude for 100%, 10%, 1%, and 0.1% of "measured" data points (selected via random sampling) are shown, with uncertainty estimates pictured beneath each reconstruction.

The differences between choice of scalarizers are illustrated in **Figure 4** and include the $x$- and $y$- components of CoM, the CoM angle, and the virtual annular bright field (ABF) signal. While physical quantities derived from electric field such as charge density and electric potential can be calculated when provided a complete data set, they cannot be computed on a *per pixel* basis since the details from neighboring pixels are required – for example, to compute gradients. For this reason, charge density and potential cannot be used in active learning experiments in which single pixels are acquired and appended to the training model. If this were desired, a single "measurement" would need to consist of a minimum of five pixels in order to compute the in-plane gradients; here, only quantities derived from single point measurements were considered. Virtually any scalar derived directly from the diffraction pattern can be used as the experimental directive to be optimized. We focus on those related to center of mass since this relates to the physical deflection of the beam and therefore atomic electric fields, but we also show the utilization of phase information - as it relates to the HAADF structure - from the ABF as a cost scalarizer.

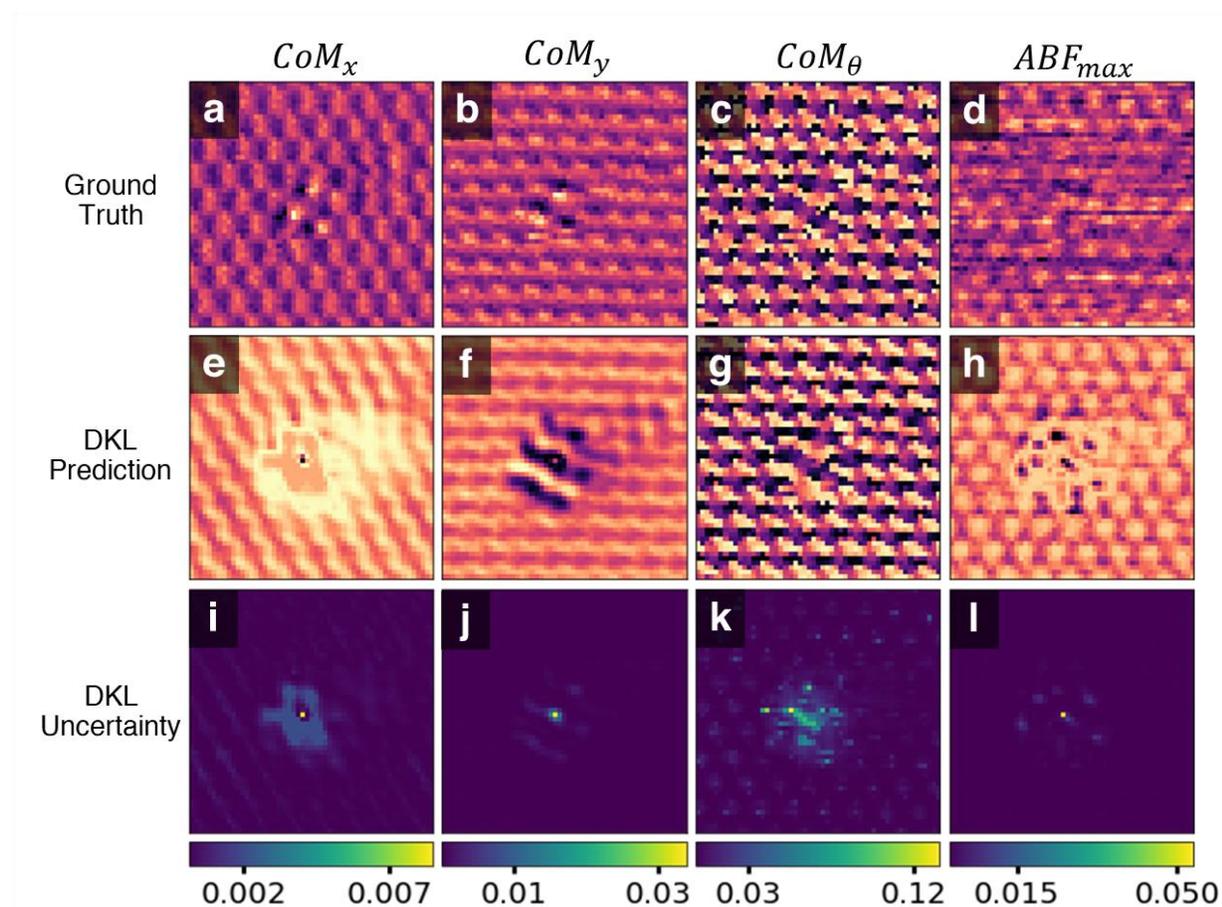

**Figure 4**. DKL reconstruction driven by different cost scalarizers. Center of mass quantities and virtual ABF used as scalarizers. Ground truths **a-d**, DKL predictions **e-h**, and uncertainties **i-l** are shown for fixed window size of 12 pixels. All reconstructions shown using complete set of data points and identical number of training steps.

Compared to using the $|CoM|$, other scalarizers of each diffraction pattern reconstruct their respective scalar functions all rather reasonably. The decision on which scalarizer to choose for optimal searching of physics is in a sense ill-posed; relationships between structure and each scalar quantity may – and likely do – exist; hence, there is no "best" choice. Interestingly, the ABF reconstruction shows a drastic improvement compared to its ground truth that contains a considerable degree of noise – this is in part due to the low-dimensional representation in latent space which naturally rejects noise. In all cases the central part of the image is clearly disrupted from the otherwise periodic patterns due to the dopant atoms, where the highest uncertainty also exists. The fields from these heavier atoms should indeed cause a stronger deflection of the beam due to their higher atomic mass and thus the presence of more bound charge.

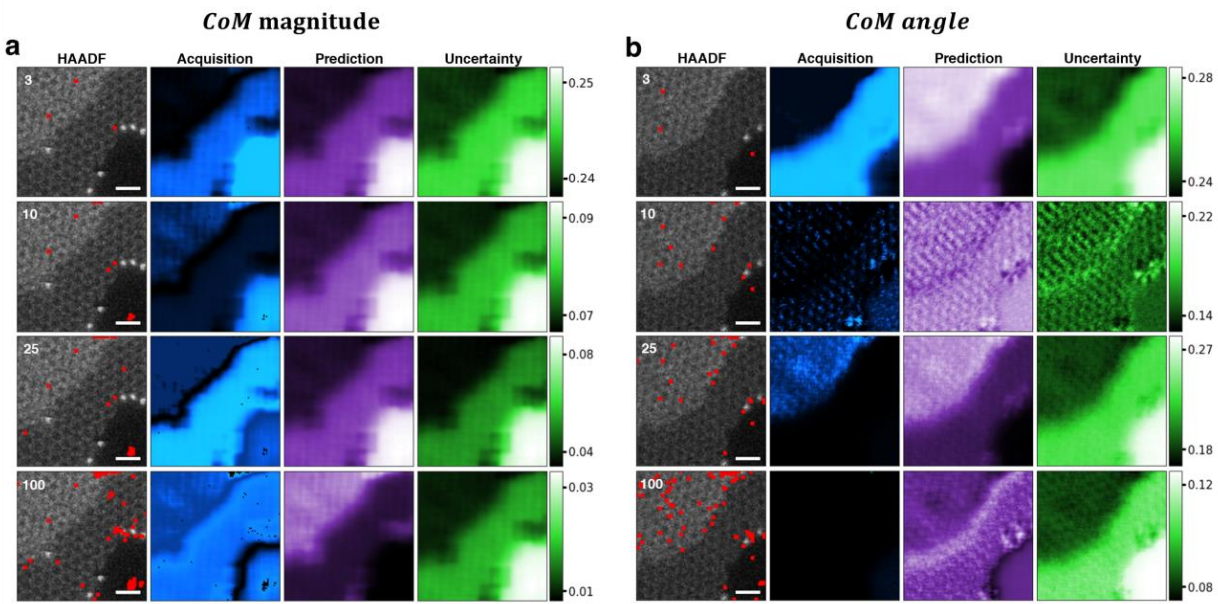

**Figure 5**. Experimental implementation of 4D-STEM DKL on the NION UltraSTEM100 microscope. Autonomous experiments are based on CoM magnitude (**a**) and CoM angle (**b**) scalarizers. HAADFs, acquisition functions, DKL predictions, and DKL uncertainties are shown at selected steps throughout the experiment. Red points on HAADF indicate visited points. AE in (a) and (b) are from the same region but acquired sequentially therefore are not identical. All scalebars 1 nm.

**Active Learning: DPC**

Finally, we deploy the DKL workflow with 4D-STEM on the operational microscope on defect-containing twisted bilayer graphene. We emphasize the fact that the model has not been pre-trained and is in fact trained on the fly and therefore can be applied to practically any material system. Using the CoM magnitude as a scalarizer, we autonomously search for regions in the HAADF that cause the strongest deflection of the beam, i.e., the strongest relative electric fields. While the CoM angle is simply the angular component of the CoM vector, it nevertheless reveals interesting phenomena in the predictions and therefore is also shown for comparison. In the live setting, the full HAADF-STEM image is acquired before collecting any 4D data such that DKL has access to the entire structural image space - as with the previous examples. Each automated experiment is performed using the same region but is not identical because the beam has induced defect formation between experiments. Despite operating at 70 kV which is below the knock-on threshold for graphene, defects and edges require substantially less energy to disrupt their bonds. Note that one could just as easily use any accelerating voltage for the automated experiments – here 70 kV was chosen only due to ideal alignment conditions at that time. For this reason, we may consider graphene that contains defects as beam-sensitive, which strengthens the motivation toward intelligent collection of 4D-STEM data. We show the AE process including acquisition functions, predictions, and uncertainties at selected steps in **Figure 5** for two different scalarizers and bring attention to several points.

First, the fact that relative field strengths are predicted reasonably well, albeit with high uncertainty, with only a handful of measurements is striking. The field strength should in principle be strongest surrounding dopants or within thicker material where more charge is accumulated; we can therefore conclude that the DKL has succeeded in learning a relationship between electric field and structure. This implies that this relationship can be established very quickly and only a small number of data points are needed to do so – indeed, even the 100 data points that were measured over the entire experiment account for 1% of the total data space. Second, in general the highest uncertainties exist within the holes (vacuum) for both scalarizers. This is expected behavior, and the DKL attempts to minimize the uncertainty there by biasing subsequent measurements to be collected from nearby or within the holes, which is observed in several of the acquisition functions by the brighter contrast near or in holes. Finally, the different pathways arising from use of different scalarizers are dramatic. The CoM magnitude is directly related to the in-plane electric field strength of the material - the early DKL predictions in Figure 5(a) assert that this should occur most strongly in the hole; at the same time however, the uncertainty is greatest here.

In DKL, successive measurements are based on either maximum prediction, maximum uncertainty, or more commonly a combination of both. The latter is known as balancing exploration and exploitation, and this is exactly what is observed – the uncertainty is slowly reduced by continually visiting regions of high uncertainty (within the hole) until which point the model learns that despite the presence of a hole, the field strength, should in reality, be stronger surrounding dopant atoms or on bilayer regions. When the CoM angle is instead used in Figure 5(b), the predictions and exploration pathway start to take on a different form, and the former even exhibits atomic contrast. Near the final step, the prediction shows features surrounding dopant atoms as well as a distinct boundary between bilayer and single layer graphene, all of which is found by considering the direction of the CoM shift instead of its magnitude. This counterintuitive result shows the power of DKL-based automated experiments and begins to uncover phenomena that are not observed in ground truth data nor would necessarily be expected. The key lies in the combined knowledge of the structural data with the scalar quantity derived from the diffraction data – these are almost always considered separately but rarely together.

**Active Learning: Strain**

With this in mind, we continue to explore the 4D space again but now we reduce our convergence angle to access crystallographic relationships. Strain and other crystal details may be extracted from nanobeam electron diffraction (NBED) by selecting a smaller convergence angle. This approach results in decreased spatial resolution and a corresponding loss of atomic contrast. Computation of strain from 4D-STEM data typically includes selection of a reference region that is expected to display minimal strain, e.g., the center of a nanoparticle. In the case where pure exploration is necessary, a reference region does not exist. Therefore, more relaxed conditions may be used; particularly we examine the exploration routes where the NBED patterns are scalarized to the average of Bragg disc centers from the center of the central beam, where only the first order diffraction discs are considered. **Figure 6** shows the ground truth extracted from the 4D NBED set and DKL pathway for maximizing Bragg distances. For a fair comparison, the ground truth here is not the conventional strain calculation, but instead is a measure of average, relative changes in lattice parameters which is momentum agnostic, as the identical scalarizer has been used in the DKL experiment. While the uncertainty in the DKL measurements remains fairly-high, the acquisition function prefers to measure near boundaries, tracing a path following the edges of graphene but mostly in vacuum. In general, it is expected that any vacuum interface displays a higher relative strain than internal structure and again DKL appears to have learned this without prior knowledge. Perhaps counterintuitively, several vacuum measurements are acquired; however, a very high level of uncertainty exists there due to the absence of diffraction. While attempting to maximize the scalarizer function, DKL simultaneously seeks to reduce uncertainty, hence often probing vacuum.

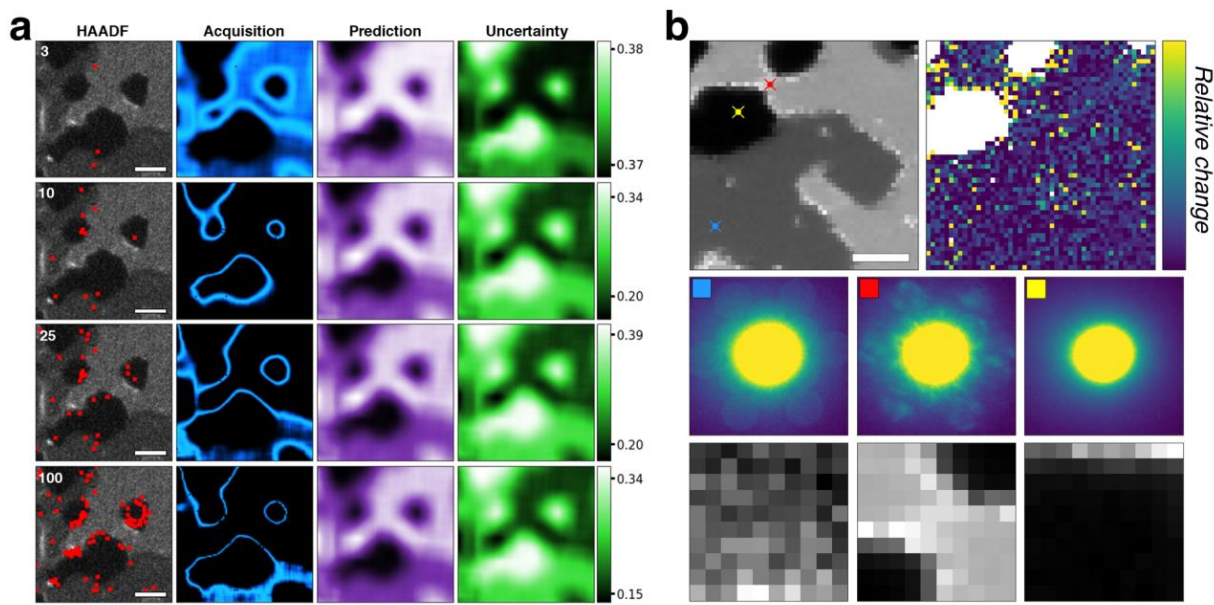

**Figure 6**. Ground truth and DKL pathway using NBED data from twisted bilayer graphene. DKL pathway, acquisition function, prediction, and uncertainty shown at indicated steps throughout active learning in **a**. Bragg disc centers are detected, with the maximum average of peak distances is used as the scalarizer. Only those corresponding to first order diffraction are considered. Ground truth HAADF and identical scalarized quantity shown in **b**, with selected diffraction patterns and matching image patches shown. Vacuum regions are masked in the ground truth. Scalebars 5 nm.

To illustrate applicability for the lesser-known systems (and hence minimize the inherent biases inevitable for well-studied materials), we deploy 4D-STEM DKL to a different material – $MnPS_3$, a layered van der Waal 2D material that is sensitive to the electron beam. $MnPS_3$ was recently investigated in the STEM and with DKL using EELS[22,27] but along with other metal phosphate trichalcogenides, it is still rather unexplored. The near-hexagonally symmetric features observed in the HAADF are a result of one or more layers of $MnPS_3$ contracting relative to its normal lattice, due to the formation of sulfur vacancies from the electron beam. To observe the interference clearly, a defocus value of $-40\ nm$ is used, hence atomic contrast is visible in the CBED patterns but not the HAADF. For comparison, an in-focus HAADF is shown adjacently to show the origins of the interference. Performing a routine 4D-STEM data collection on the $MnPS_3$ would impart too much dose to the specimen, significantly altering the material during acquisition. It is for this reason that we instead aim to explore this unusual ordering mechanism through the automated DKL experiment, which significantly reduces the total dose by simply visiting only a small number of total points. Since only a small number of points are ever visited, e.g., 3%, effectively the dose can be reduced by a factor of about 30. This factor can be improved even further if the DKL is pre-trained, reducing - possibly to zero - the number of acquisitions needed for training – the only requirement that does not vanish is acquisition of the HAADF-STEM image at the start of the experiment, but that generally only takes on the order of tens of milliseconds.

In **Figure 7** we show that using the same 4D-STEM CoM criterion as before, the DKL model begins to learn the relationship these ordered structures have with electric field strengths. The model predicts very quickly that the strongest CoM signal should be correlated directly to the higher intensity regions observed in the HAADF, which corresponds to the constructive interference of multiple layers. The destructively interfering regions however exhibit a weaker field strength, and the model here contains higher levels of uncertainty, which is why many of the measured points come from regions surrounding the brighter circular regions. This experiment exemplifies how DKL experiments in 4D-STEM can be used to investigate dose sensitive specimens, avoid preconceived biases in selecting regions of interest, and learn relationships between the structure and local diffraction.

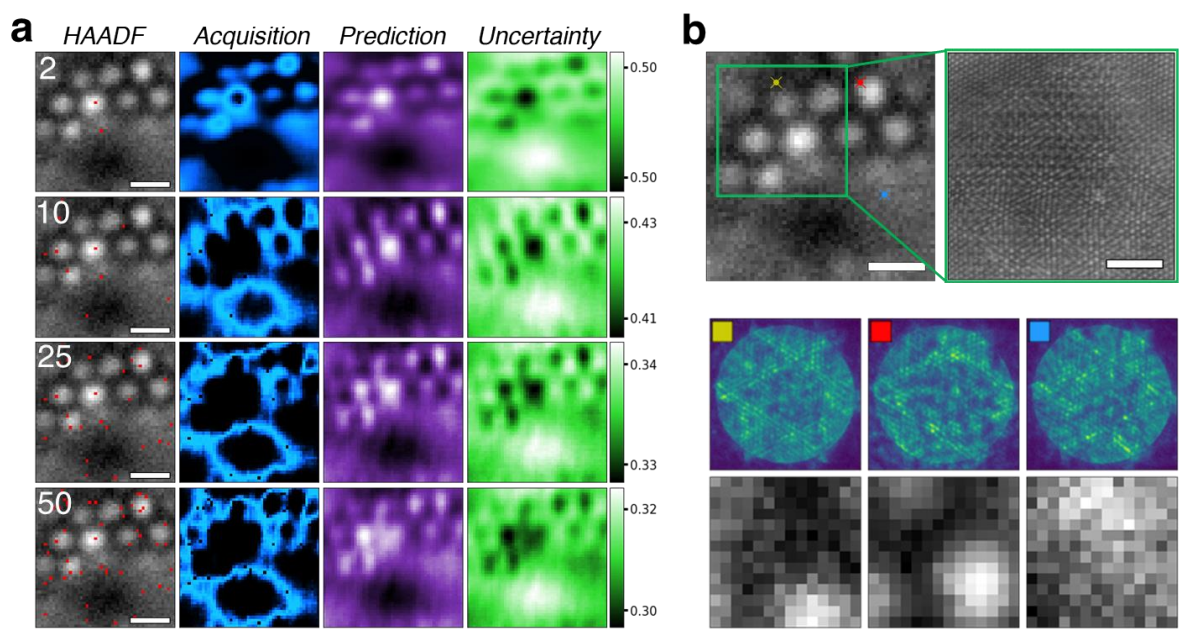

**Figure 7**. Active learning with DPC CoM scalarizer on MnPS$_3$, using a defocus value of $-40\ nm$. DKL pathway, acquisition function, prediction, and uncertainty shown at indicated steps throughout active learning in **a**. Periodic interference is a result of contracting layers relative to one another due to generation of sulfur vacancies. Ground truth HAADF and in-focus HAADF shown in **b**, with selected diffraction patterns and corresponding image patches. Scalebar 5 nm and 2 nm.

CONCLUSIONS

Finally, we explore computational requirements and possible edge computing needs. Here, the DKL model is re-trained with each new measurement, which takes about 1-2 s with an on-board graphics processing unit (GPU); here an NVIDIA Quadro K2200 was used. If a GPU is inaccessible, using the CPU can cause this time to increase by about one order of magnitude.

Considering that a single diffraction pattern with a CMOS detector requires about a 10 ms acquisition time, the model training time is absolutely the bottleneck when attempting to minimize overall experiment time. While we do not necessarily claim that automated experiments enabled by DKL are intended to save time, we do point out the previous observation that structure-property relationships can be learned with a few number of exploration steps – or learned in an offline setting. In other words, DKL can in principle be used to save time for intelligent acquisitions.

To summarize, we have developed the active learning methods for the 4D-STEM imaging. Using the deep kernel learning, we enable discovery of the behaviors of interest driven by the internal fields, and demonstrate this approach for twisted bilayer graphene and mesoscale-ordered patterns in the $MnPS_3$. We also note there are opportunities to increase the rate of the training by using pre-acquired data to train invariant variational autoencoders,[39] and then use the pre-trained weights to initialize the DCNN part of the DKL. Similarly, the interventional strategies utilizing the knowledge derivable from physical deconvolution of the 4D-STEM patterns can be used to structure the latent space via conditioning or bootstrapping strategies, hence allowing for more directed physical search. This approach can be extended to physics-based discovery of novel physical and chemical phenomena in strongly correlated and quantum materials.

METHODS/EXPERIMENTAL

**Materials**

*Graphene*
Graphene was grown on Cu foil using atmospheric pressure chemical vapor deposition (AP-CVD)[40] Poly (methyl methacrylate) (PMMA) was spin coated on the surface to form a mechanical stabilizer. The Cu foil was etched away in a bath of deionized water (DI) and ammonium persulfate. The PMMA/graphene stack was rinsed in DI water and caught on a Protochips heater chip. The chip was then baked at 150 C on a hot plate for 15 mins before being immersed in acetone for 15 mins to dissolve the PMMA. Upon removal from the acetone, the chip was dipped in isopropyl alcohol and allowed to dry.

*$MnPS_3$*
$MnPS_3$ single crystals were synthesized through the chemical vapor transport (CVT) method. Manganese powder (Alfa Aesar 99.95%), phosphorus powder (Alfa Aesar 99.995%) and Sulfur chunks (Puratronic 99.9995%) of a ratio 1: 1: 3.1 were thoroughly mixed and ground under argon atmosphere. The mixture was then pressed into pellet and sealed in a fused tube under vacuum. Polycrystalline powder was obtained after annealing the ampoule in a muffle furnace for a week at 730 °C.
Two grams of annealed powder were transfer into a new fused tube. Dehydrated iodine was added as a transport agent, then the tube was connected to a vacuum station and sealed at a proper length. Large single crystals could be harvested at the cold end of the tube after annealing at 700C for six days in a tube furnace.
Electron transparent flakes were isolated by mechanical exfoliation and were directly transferred to Au Quantifoil TEM grids, avoiding any transfer step to silicon.

**Microscopy**

Electron microscopy was carried out using a NION 5$^{th}$ order aberration corrected scanning transmission electron microscope, with an accelerating voltage of 70 kV and a nominal probe current of 30 pA. A semiconvergence angle of 30 mrad was selected for CoM and virtual aperture experiments, while 5 mrad was used for strain experiments. A fast CMOS detector was needed for strain experiments with weakly scattering single layer graphene, but a traditional CCD was still viable for experiments at a semiconvergence angle of 30 mrad - even in graphene. Pixel dwell times between 2 and 250 ms were used, depending on type of experiment. All specimens were heated to 160° C in high vacuum overnight prior to inserting into the microscope.

**Implementation of Automated Experiments**

Automated experiments were made possible with the NION Swift control interface[41] and were entirely implemented in Python directly on the microscope user PC. Swift allows direct access to many microscope controls such as probe position and measurement acquisition via Python scripting in real time; therefore, automated experiments can be conducted without the need of any external hardware. At the same time, model training and prediction can also occur directly in the Python Swift environment using the on-board GPU (or even CPU), hence there is no need to transfer data off the microscope. All available control commands are documented on the Nion Swift Github / webpage.[44] The DKL workflow was realized via AtomAI.[42] While GPU utilization was not fundamentally necessary, it helped to decrease overall model training times by at least one order of magnitude

**Acquisition function in DKL**

Throughout this paper, we used the expected improvement (EI) acquisition function, which tells a likelihood of the highest improvement over the current "best measurement" and is defined as[43]

$$\alpha_{EI} = (\mu(\mathbf{x}) - y^+ - \xi)\Phi\left(\frac{\mu(\mathbf{x}) - y^+ - \xi}{\Sigma(\mathbf{x})}\right) + \Sigma(\mathbf{x})\phi\left(\frac{\mu(\mathbf{x}) - y^+ - \xi}{\Sigma(\mathbf{x})}\right)$$

where $\Phi$ is a standard normal cumulative distribution function, $y^+$ is the best predicted value, $\phi$ is the standard normal probability density function, and $\xi$ balances the exploration and exploitation (here it was set to 0.01).

**Data availability**

The data used for analysis as well as additional materials are available through the Jupyter notebook located at: https://github.com/kevinroccapriore/AE-DKL-4DSTEM

**Conflict of Interest**

The authors declare no conflict of interest.

ASSOCIATED CONTENT

AUTHOR INFORMATION

**Corresponding Authors**

Kevin M. Roccapriore: roccapriorkm@ornl.gov

Sergei V. Kalinin: sergei2@utk.edu

ACKNOWLEDGMENT

This research is sponsored by the INTERSECT Initiative as part of the Laboratory Directed Research and Development Program of Oak Ridge National Laboratory, managed by UT-Battelle, LLC for the US Department of Energy under contract DE-AC05-00OR22725 (K.M.R). This effort was based upon work supported by the U.S. Department of Energy (DOE), Office of Science, Basic Energy Sciences (BES), Materials Sciences and Engineering Division (M.O., O.D., S.V.K.). Work was partially supported (M.Z.) and conducted using resources supported by Oak Ridge National Laboratory's Center for Nanophase Materials Sciences (CNMS), which is a U.S. DOE Office of Science User Facility. Electron microscopy was performed using instrumentation within ORNL's Materials Characterization Core provided by UT-Battelle, LLC, under Contract No. DE-AC05- 00OR22725 with the DOE and sponsored by the Laboratory Directed Research and Development Program of Oak Ridge National Laboratory, managed by UT-Battelle, LLC, for the U.S. Department of Energy. The authors greatly appreciate the $MnPS_3$ samples provided by Nan Huang and David G. Mandrus from the University of Tennessee.